\documentstyle[epsfig]{article} 

\begin{document} 
\title{Nonlinear projective filtering in a data stream}
\author{Thomas Schreiber and Marcus Richter \\
      Physics Department, University of Wuppertal,\\ 
      D--42097 Wuppertal, Germany}
\maketitle
\begin{abstract}
  We introduce a modified algorithm to perform nonlinear filtering of a time
  series by locally linear phase space projections. Unlike previous
  implementations, the algorithm can be used not only for {\sl a posteriori}
  processing but includes the possibility to perform real time filtering in a
  data stream. The data base that represents the phase space structure
  generated by the data is updated dynamically. This also allows filtering of
  non-stationary signals and dynamic parameter adjustment. We discuss exemplary
  applications, including the real time extraction of the fetal
  electrocardiogram from abdominal  recordings.\\[0.5cm]
  PACS: 05.45.+b\\[0.5cm]
\end{abstract}

Nonlinear projective filtering of time series is one of the most practical
outcomes of the theory of nonlinear dynamical systems as applied to real word
observations.  Signals with nonlinear dynamical correlations are often not
handled properly by spectral processing methods. Nonlinear dynamical systems
exhibiting deterministic chaos have been proposed as an alternative paradigm
for the study of such complex sequences. However, for the derivation of time
series methods within this framework, heavy use had to be made of the
theoretical properties of deterministically chaotic systems. This seems to
severely restrict the range of systems they may be applied to -- after all,
very few real world phenomena are adequately described by a purely
deterministic time evolution. Also nonlinear filtering procedures were
designed to exploit the specific structure generated by deterministic dynamics
in phase space.  It turns out, however, that careful use of these algorithms
can give excellent results also in situations where pure determinism is
violated. The reason is that when represented in a low-dimensional phase space,
also non-deterministic systems may exhibit structures suitable for filtering
purposes.  One example is the human electro-cardiogram (ECG), a signal that
shows features of nonlinear determinism but also essential stochastic
components.

An introduction to deterministic chaos and dynamical systems as a framework
for the analysis of time series recordings can be found in the monographs
Refs.~\cite{Ablabla,KantzSchreiber}. Review articles on nonlinear filtering
algorithms containing further references to the original literature are
Refs.~\cite{ks,Davies}. How and why the ECG can be processed with nonlinear
phase space filters is discussed in~\cite{marcus}.

Apart from these fundamental considerations, there are also some practical
issues that have so far hampered widespread use of nonlinear filters.  All of
the methods that have been proposed in the literature are formulated as {\sl a
posteriori} filters. The whole signal has to be available before a cleaned
version can be computed. The actual computation is invariably quite
computer time intensive. One class of methods uses a global nonlinear function
to represent an approximation to the dynamics. This function has to be
determined by a delicate fitting procedure and the actual filtering scheme (for
example Ref.~\cite{davies}) consists of an iterative minimisation procedure in
a high-dimensional space. The other class of algorithms approximates the
dynamics or geometry in phase space by locally linear mappings. Here, covering
phase space with small neighbourhoods is the most time consuming step, along
with the need to solve a least squares or eigenvalue problem in each of these
neighbourhoods. With fairly low-dimensional signals and small noise levels,
fast neighbour search algorithms (see~\cite{neigh} for an overview) are very
helpful in this regard.  In any case, the posterior nature and the
computational effort has been one of the major reasons why nonlinear phase
space filters haven't seen more widespread use. The purpose of this article is
to introduce modifications to a locally projective noise reduction scheme that
make its use in real time signal processing feasible.

In brief, we implement three main modifications to the locally projective noise
reduction scheme that has been introduced in Ref.~\cite{on}. Exactly the same
strategy can be applied to modify the algorithms by Sauer~\cite{sauer} or that
by Cawley and Hsu~\cite{chs}. (1) The data base of local neighbours that is
needed to approximate the dynamics is restricted to points in the past, thereby
rendering the filter causal. Of course, at the beginning no data base is
available and noise reduction gradually becomes more effective as new points
are collected. As a side effect, the curvature correction proposed
in~\cite{sauer} and discussed in~\cite{KantzSchreiber} can be carried out
during the first sweep through the data. This will be explained below.  (2) The
number of neighbours required to carry out the correction for each point is
limited to a number that is just sufficient for statistical stability of the
local linear fits. (3) The last and most severe modification uses the fact that
the dynamics is supposed to vary smoothly in phase space. Therefore, instead of
determining the locally linearised dynamics at each point, the linear
approximation is stored only for a collection of representative points which
densely cover the observed set of points.  Consequently, the local linear
problem has to be solved only for points which are about to become a
representative.

The algorithm we present in this paper is based on the noise reduction scheme
introduced in Ref.~\cite{on}. An alternative derivation can be found in
Ref.~\cite{processing}. We refer the reader to these references as well as the
monograph Ref.~\cite{KantzSchreiber} for the motivation of the approach by the
theory of deterministic dynamical systems. In a more general context, a scalar
time series $\{s_n\}, n=1,\ldots,N$ can be unfolded in a multi-dimensional
effective phase space using time delay coordinates ${\bf
s}_n=(s_{n-(m-1)\tau},\ldots,s_n)$ ($\tau$ is a delay time). If $\{s_n\}$ is a
scalar observation of a deterministic dynamical system, it can be shown under
certain genericity conditions~\cite{takens,embed} that the reconstructed point
set is a one-to-one image of the original attractor of the dynamical system.
We will not explicitly assume here that there is such an underlying
deterministic system. Nevertheless, general serial dependencies among the
$\{s_n\}$ will cause the delay vectors $\{{\bf s}_n\}$ to fill the available
$m$-dimensional space in an inhomogeneous way. Linearly correlated Gaussian
random variates will for example be distributed according to an anisotropic
multivariate Gaussian distribution. Linear geometric filtering in phase space
seeks to identify the principal directions of this distribution and project
onto them. This concept is implemented by the singular systems approach, see
Refs.~\cite{svd,svd2,vg} and many others. The present algorithm can be seen as
a nonlinear generalisation of this approach which takes into account that
nonlinear signals will form curved structures in delay space. In particular,
noisy deterministic signals form smeared-out lower-dimensional
manifolds. Nonlinear phase space filtering seeks to identify such structures
and project unto them in order to reduce noise.

Thus, conceptually, the noise reduction algorithm consists of three main steps.
(1) Find a low-dimensional approximation to the ``attractor'' described by the
trajectory $\{{\bf s}_n \}$. (2) Project each point ${\bf s}_n$ on the
trajectory orthogonally onto the approximation to the attractor to produce a
cleaned vector $\hat{\bf s}_n$ (3) Convert the sequence of cleaned vectors
$\hat{\bf s}_n$ back into the scalar time domain to produce a cleaned time
series $\hat{s}_n$.

\begin{figure}
\centerline{
\begin{picture}(0,0)%
\epsfig{file=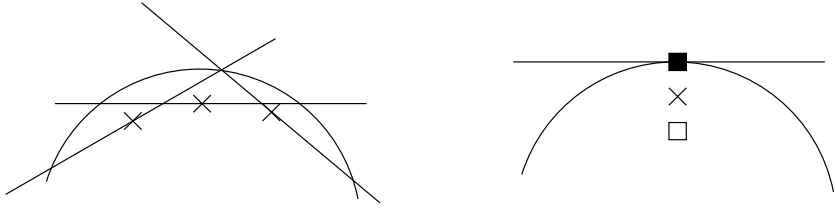}%
\end{picture}%
\setlength{\unitlength}{3947sp}%
\begingroup\makeatletter\ifx\SetFigFont\undefined%
\gdef\SetFigFont#1#2#3#4#5{%
  \reset@font\fontsize{#1}{#2pt}%
  \fontfamily{#3}\fontseries{#4}\fontshape{#5}%
  \selectfont}%
\fi\endgroup%
\begin{picture}(5219,1282)(95,-2789)
\end{picture}
}
\caption[]{Local linear approximation to a one-dimensional
   curve. Left: approximations are not tangents but secants and all the
   centres-of-mass (crosses) of different neighbourhoods are shifted inward
   with respect to the curvature. Right: a tangent approximation is obtained by
   shifting the centre-of-mass outward with respect to the curvature. The open
   square denotes the average of the centres-of-mass of adjacent
   neighbourhoods, the filled square is the corrected centre-of-mass.
   \label{fig:tangent}}
\end{figure}

The low-dimensional approximation to the point set can be constructed locally
in delay coordinate space using a procedure that is very similar to a local
version of principal component analysis.  In order to define a neighbourhood in
phase space around a point ${\bf s}_n$, find all of the points that are within
a distance $\epsilon$ of ${\bf s}_n$.  A set of nearby points can be defined as
\begin{equation}
   {\cal U}^{(n)}_{\Delta n} = \{{\bf s}_n':\quad n-\Delta n \le n'\le n, \quad
           \| {\bf s}_n'-{\bf s}_n\| < \epsilon\}
\end{equation}
where $\|{\bf s}_n' -{\bf s}_n \|$ is the phase space distance between ${\bf
s}_n'$ and ${\bf s}_n$. (We use the max norm to measure distances.) In this
definition we already included two modifications with respect
to~\cite{on}. First, only points in the past $(n'\le n)$ are considered and
second, neighbours are only considered that have been observed no longer than
$\Delta n$ time steps ago. The neighbourhood ${\cal U}^{(n)}$ which is actually
used for noise reduction is determined by finding the largest $\Delta n<n$ for
which the number $|{\cal U}^{(n)}|$ of vectors in ${\cal U}^{(n)}$, does not
exceed a specified $U_{\mbox{max}}$.

The local structure of the point set is approximated by a linear subspace
formed essentially by local principal components. The most straightforward
implementation is to compute the local centre-of-mass
\begin{equation}
   \langle{\bf s}\rangle^{(n)} = |{\cal U}^{(n)}|^{-1}
      \sum_{{\bf s}_{n'}\in {\cal U}^{(n)}} {\bf s}_{n'}
\end{equation} 
where $\sum_{{\bf s}_{n'}\in {\cal U}^{(n)}} $ means summation over all $|{\cal
U}^{(n)}|$ vectors in ${\cal U}^{(n)}$.  Then the principal components can be
calculated around this point. In that case, however, the linear subspace is not
tangent to the curved manifold but rather intersects with it, as illustrated in
Fig.~\ref{fig:tangent}.  Therefore, it is preferable to use a corrected
centre-of-mass $\overline{\bf s}^{(n)}$ given by
\begin{equation}\label{eq:cm}
   \overline{\bf s}^{(n)} = 2\,\langle{\bf s}\rangle^{(n)}
      - |{\cal U}^{(n)}|^{-1} \sum_{{\bf s}_{n'}\in {\cal U}^{(n)}} 
      \langle{\bf s}\rangle^{(n')} 
\end{equation} 
In the original implementation~\cite{on}, neighbourhoods could contain future
points. Thus, Eq.(\ref{eq:cm}) could only be evaluated after one complete sweep
through the data in which all the local centres-of-mass $\langle{\bf
s}\rangle^{(n)}$ were determined. In order to compute the tangent points
$\overline{\bf s}^{(n)}$, a second sweep through the data set was necessary.
In this modified implementation, the centre-of-mass vectors $\langle{\bf
s}\rangle^{(n)}$ are stored when the point with index $n$ is processed. Thus
all these vectors are available for points with index $n'\le n$. Thus,
$\overline{\bf s}^{(n)}$ can be formed immediately.

Now, the local weighted covariance matrix
\begin{equation}
   C_{ij}^{(n)} = \sum_{{\bf s}_{n'}\in {\cal U}^{(n)}} 
       [{\bf R}({\bf s}_{n'}-\overline{\bf s}^{(n)})]_i 
       [{\bf R}({\bf s}_{n'}-\overline{\bf s}^{(n)})]_j 
\end{equation}
is computed, where $[\cdot]_i$ denotes the $i$--th component of the vector in
brackets.  As discussed in Ref.~\cite{on}, the weight matrix $\bf R$ is chosen
to be diagonal with $R_{11}$ and $R_{mm}$ large and all other diagonal entries
$R_{ii}=1$.  Now determine the orthonormal eigenvectors ${\bf c}_q$ and
eigenvalues of $C_{ij}^{(n)}$ using standard matrix techniques. A
$Q$-dimensional manifold is then locally approximated by those $Q$ eigenvectors
with the largest eigenvalues. The projected vector $\hat{\bf s}_n$ is then
given by:
\begin{equation} 
   \hat{\bf s}_n =\overline{\bf s}^{(n)} 
   + {\bf R}^{-1} \sum_{q=1}^Q {\bf c}^q 
      [{\bf c}^q \cdot{\bf R}({\bf s}_n-\overline{\bf s}^{(n)})]
\,.\end{equation}
In order to translate $\{ {\bf s}_n \}$ back into a scalar signal, we note that
each scalar measurement $s_n$ appears as a component in $m$ embedding vectors,
${\bf s}_n,\ldots,{\bf s}_{n+m-1}$. The corrected scalar time series values
$\hat s_n$ are thus obtained by averaging the corresponding components of
$\hat{\bf s}_n,\ldots,\hat{\bf s}_{n+m-1}$. 

For real time application, this means that the corrected value $\hat s_n$
cannot be available before $s_{n+m-1}$ has been measured and processed.
Usually, however, this delay window is a very short time, at least compared to
the duration of the recording, and the procedure can be regarded as
effectively on-line as long as the computations necessary to obtain $\hat s_n$
can be carried out fast enough. 

So far, we have turned the procedure into a causal filter by restricting
neighbour search to points defined by measurements made in the past. By further
limiting the number of neighbours searched for, we have sped up the formation
of the local covariance matrices considerably. Still, as the algorithm stands,
we have to solve an $(m\times m)$ eigenvalue problem for each point that is to
be processed. A large fraction of this work can be avoided on the base of the
assumption that the local linear structure changes smoothly over phase
space. By making local linear approximations we have already assumed smoothness
of the underlying manifold in the $C_1$ sense. In most physical systems, the
additional assumption of $C_2$ smoothness is not less justified. We cannot,
however expect that the vectors which span the principal directions vary slowly
from point to point. The reason is that often some eigenvalues are nearly
degenerate and change indices from point to point when they are ordered by
their magnitude.  We thus refrain from interpolating principal components
between phase space points.  Instead, we choose a length scale $h$ in phase
space which is small enough such that the linear subspaces spanned by the local
principal components can be regarded as effectively the same.  Now we
successively build up a data base of representative points for which the local
points of tangency $\overline{\bf s}^{(n')}$, and the local principal
directions ${\bf c}^q,\quad q=1,\ldots,Q$ have been determined already. For
each new point ${\bf s}_n$ that is to be processed, we go through this
collection of points to determine whether a representative is available closer
than $h$. In that case, we use the stored tangent point and principal
directions of the representative in order to perform the projections. If not, a
neighbourhood is formed around ${\bf s}_n$ in which the eigenvalue problem is
solved. The point ${\bf s}_n$ is then included in the list of
representatives. By this procedure, all parts of the available space that is
visited by observations are covered by representatives with a maximal distance
$h$. If the dynamics and thus the geometry in phase space undergoes some change
during the recording, new areas are visited which are automatically covered by
new representatives.  It is also possible to delete representatives which are
older than a given time span.  A realisation of the algorithm that implements
all the modifications discussed in this paper is publicly available as part of
the TISEAN software project~\cite{tisean}.

\begin{figure}
\centerline{\hspace*{0.5cm}%
\setlength{\unitlength}{0.1bp}
\begin{picture}(2160,1728)(0,0)
\includegraphics{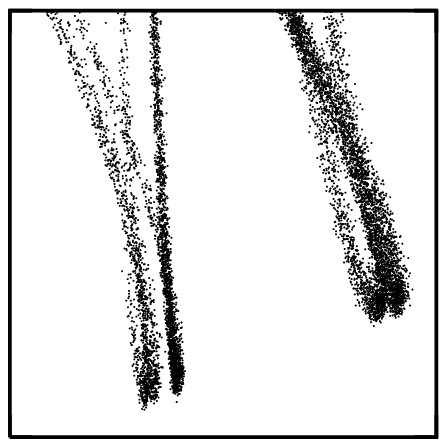}
\put(1164,150){\makebox(0,0){$x(t-\mbox{11 ms})$}}
\put(200,1014){%
\makebox(0,0)[b]{\shortstack{$x(t)$}}%
}
\put(1778,300){\makebox(0,0){3000}}
\put(550,300){\makebox(0,0){1500}}
\put(500,1628){\makebox(0,0)[r]{-2500}}
\put(500,400){\makebox(0,0)[r]{-4000}}
\end{picture}%
\hspace*{-1.5cm}%
\setlength{\unitlength}{0.1bp}
\begin{picture}(2160,1728)(0,0)
\includegraphics{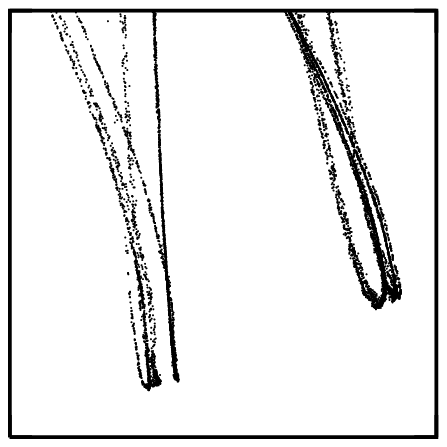}
\put(1164,150){\makebox(0,0){$x(t-\mbox{11 ms})$}}
\put(200,1014){%
\makebox(0,0)[b]{\shortstack{$x(t)$}}%
}
\put(1778,300){\makebox(0,0){3000}}
\put(550,300){\makebox(0,0){1500}}
\put(500,1628){\makebox(0,0)[r]{-2500}}
\put(500,400){\makebox(0,0)[r]{-4000}}
\end{picture}
}
\caption[]{\label{fig:raser}Result of nonlinear filtering of an NMR laser time
  series. An enlargement of about one quarter of the total linear extent of
  the attractor is shown.}
\end{figure}

As a first example, we show in Fig.~\ref{fig:raser} the result of applying the
described procedure to a data set from an NMR laser experiment~\cite{raser}.
The same data has also been used in Ref.~\cite{buzug}. The laser is
periodically driven and once per driving cycle the envelope of the laser output
is recorded. The resulting sampling rate is 91~Hz. At this rate, the modified
nonlinear noise reduction scheme can be easily carried out in real time on a
Pentium~II processor at 200~MHz. Since further iterations can be carried out
after the time corresponding to one embedding window without interfering with
the previous steps, we could perform up to three iterations on a dual
Pentium~II workstation at 300~MHz in real time. The figure shows the result
after two iterations. Projections from $m=7$ down to $Q=2$ dimensions were
used, at least 100 neighbours were requested at $\epsilon=200$~A/D units. The
history was limited to 20000 samples, or 220~s. This fairly large data base is
needed since the initial noise level is already small (less than
2\%~\cite{buzug}) and small neighbourhoods are required to avoid curvature
artefacts. The maximal distance of representative points was chosen to be
$h=120$~A/D units.

\begin{figure}
\centerline{\hspace{1.5cm}%
\setlength{\unitlength}{0.1bp}
\begin{picture}(2160,1728)(0,0)
\includegraphics{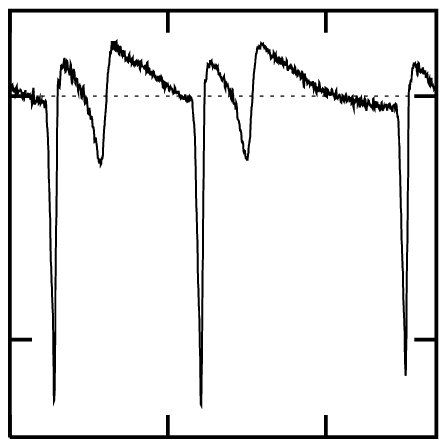}
\put(1014,150){\makebox(0,0){$t$ [s]}}
\put(219,1014){%
\makebox(0,0)[b]{\shortstack{$x(t)$ [A/D units]}}%
}
\put(1310,300){\makebox(0,0){2}}
\put(855,300){\makebox(0,0){1}}
\put(400,300){\makebox(0,0){0}}
\put(350,1382){\makebox(0,0)[r]{0}}
\put(350,681){\makebox(0,0)[r]{-1}}
\end{picture}%
\hspace*{-1.5cm}%
\setlength{\unitlength}{0.1bp}
\begin{picture}(2160,1728)(0,0)
\includegraphics{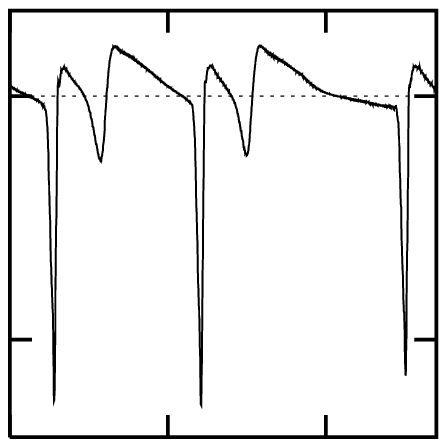}
\put(1014,150){\makebox(0,0){$t$ [s]}}
\put(219,1014){%
\makebox(0,0)[b]{\shortstack{$x(t)$ [A/D units]}}%
}
\put(1310,300){\makebox(0,0){2}}
\put(855,300){\makebox(0,0){1}}
\put(400,300){\makebox(0,0){0}}
\put(350,1382){\makebox(0,0)[r]{0}}
\put(350,681){\makebox(0,0)[r]{-1}}
\end{picture}
}
\caption[]{Result of nonlinear filtering of an MCG time series. Note that the
noise is not white. It contains for example contributions from non-cardiac
muscle activity and fluctuations in the magnetic background 
field.\label{fig:mcg}}
\end{figure}

\begin{figure}
\centerline{\hspace{1cm}%
\setlength{\unitlength}{0.1bp}
\begin{picture}(2160,1728)(0,0)
\includegraphics{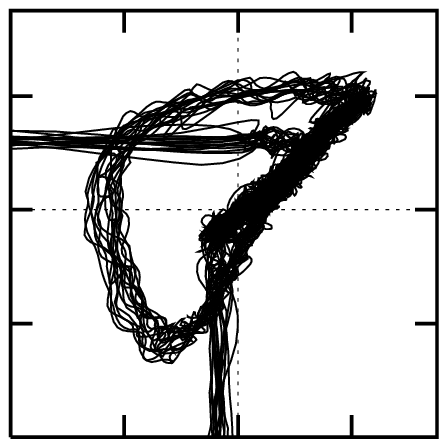}
\put(1114,150){\makebox(0,0){$x(t-\mbox{40 ms})$}}
\put(200,1014){%
\makebox(0,0)[b]{\shortstack{$x(t)$}}%
}
\put(1482,300){\makebox(0,0){0.2}}
\put(1155,300){\makebox(0,0){0}}
\put(827,300){\makebox(0,0){-0.2}}
\put(450,1382){\makebox(0,0)[r]{0.2}}
\put(450,1055){\makebox(0,0)[r]{0}}
\put(450,727){\makebox(0,0)[r]{-0.2}}
\end{picture}%
\hspace*{-1.5cm}%
\setlength{\unitlength}{0.1bp}
\begin{picture}(2160,1728)(0,0)
\includegraphics{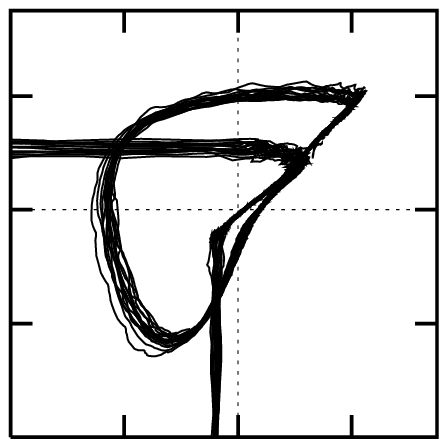}
\put(1114,150){\makebox(0,0){$x(t-\mbox{40 ms})$}}
\put(200,1014){%
\makebox(0,0)[b]{\shortstack{$x(t)$}}%
}
\put(1482,300){\makebox(0,0){0.2}}
\put(1155,300){\makebox(0,0){0}}
\put(827,300){\makebox(0,0){-0.2}}
\put(450,1382){\makebox(0,0)[r]{0.2}}
\put(450,1055){\makebox(0,0)[r]{0}}
\put(450,727){\makebox(0,0)[r]{-0.2}}
\end{picture}
}
\caption[]{Result of nonlinear filtering of an MCG time series. Same data as
   Fig.~\ref{fig:mcg} but in time delay representation (enlargement).
\label{fig:mcgd}}
\end{figure}

The actual acceleration resulting from the above modifications strongly depends
on the situation and it is difficult to give general rules and benchmarks. Let
us however study a realistic example in some detail to illustrate the main
points. In electrophysiological research, it is quite attractive to augment the
measurements of electric potentials with recordings of the magnetic field
strength. The latter penetrate intervening tissue much more efficiently. Of
particular interest are {\em magneto-encephalographic} (MEG) recordings which
allow to access regions of the brain noninvasively which cannot be monitored
electrically using surface electrodes. In cardiology, the {\em
magneto-cardiogram} (MCG) provides additional information to the traditional
{\em electrocardiogram} (ECG). A particular application is the noninvasive
monitoring of the fetal heart which is otherwise complicated by shielding of
the electric field by intervening tissue. A common problem with magnetic
recordings, however, is that the fields are rather feeble and the measurements
have to be carried out in a shielded room. Even then, noise remains a major
challenge for this experimental technique. We will demonstrate in the following
how nonlinear noise reduction could be used for continuous MCG monitoring.

We use an MCG recording of a normal human subject at rest. The data was kindly
provided by Carsten Sternickel at the University of Bonn. The sampling rate was
1000~Hz, which is quite high. For the signal processing task, any sampling rate
above about 200~Hz would be sufficient. (Below 200~Hz, the spike representing
the depolarisation of the ventricle might not be resolved properly). In order
to cover a significant fraction of one cardiac cycle by an embedding window, we
chose an embedding with delay $\tau =10$~ms in $m=10$ dimensions.
Neighbourhoods were formed with a radius of $\epsilon=0.1$ (in the uncalibrated
A/D units of the recording), about three times the noise level estimated by
visual inspection. In Figs.~\ref{fig:mcg} and~\ref{fig:mcgd}, the result of two
iterations of the noise reduction scheme is shown.  We quote computation times
for the processing of 10~s of MCG on a Pentium II processor at 300~MHz,
determined on a PC running the Linux operating system. The timing results are
summarised in Table~\ref{table}. The data base of representatives for (f) was
formed by assuming that the local linear subspaces are equivalent on length
scales of the order of 0.06 A/D units.

\begin{table}
\begin{center}
\begin{tabular}{llr}
\hline
& method & CPU time \\
\hline
 a) & all neighbours                   & 165~s \\
 b) & box assisted neighbour search    &  31~s \\ 
 c) & all neighbours in past           &  82~s \\
 d) & $n-n'<5$~s                       &  64~s \\
 e) & (d) and $U_{\mbox{max}}<200$     &  15~s \\
 f) & (e) and reuse of representatives &   2~s \\
\hline
\end{tabular}
\end{center}
\caption[]{Computation time for one iteration of the nonlinear noise reduction
  scheme, applied to 10~s of an MCG  recording sampled at 1000~Hz. See text for
  details.\label{table}} 
\end{table}

\begin{figure}
\centerline{\hspace*{-2cm}%
\setlength{\unitlength}{0.1bp}
\begin{picture}(2880,1728)(0,0)
\includegraphics{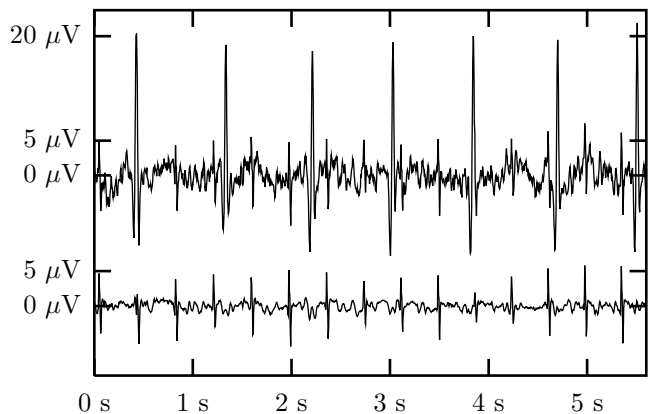}
\put(2607,150){\makebox(0,0){5 s}}
\put(2236,150){\makebox(0,0){4 s}}
\put(1864,150){\makebox(0,0){3 s}}
\put(1493,150){\makebox(0,0){2 s}}
\put(1121,150){\makebox(0,0){1 s}}
\put(750,150){\makebox(0,0){0 s}}
\put(700,644){\makebox(0,0)[r]{5 $ \mu $V}}
\put(700,512){\makebox(0,0)[r]{0 $ \mu $V}}
\put(700,1530){\makebox(0,0)[r]{20 $ \mu $V}}
\put(700,1136){\makebox(0,0)[r]{5 $ \mu $V}}
\put(700,1005){\makebox(0,0)[r]{0 $ \mu $V}}
\end{picture}
}
\caption{Fetal ECG extraction by nonlinear filtering. Upper trace: the 
  original abdominal recording. Lower trace: the extracted fetal ECG. See text
  for details.\label{fig:fekg}}
\end{figure}

Like in the case of the MCG time series demonstrated above, nonlinear phase
space filtering can be successfully applied to ECG recordings~\cite{SK}.  The
modifications of the algorithm described in the present paper allow to perform
the filtering as a real time application on common personal computers without a
noticeable impairment in signal quality compared to previous
implementations. Since this application is similar to the MCG filtering
described above, we do not need to state further details.  Instead, let us
discuss a related signal separation problem of practical relevance, the
extraction of the fetal ECG from abdominal recordings during
pregnancy~\cite{SK2,marcus2}. In order to obtain the strongest possible signal
of the fetal cardiac activity, electrodes are usually placed on the maternal
abdomen.  The upper trace of Fig.~\ref{fig:fekg} contains such an abdominal
recording. (In this particular case one abdominal and one cervical electrode
has been used; the data were kindly provided by J.~F.~Hofmeister,
Denver~\cite{recording}.)  Due to the large size of the maternal heart in
comparison to the fetal one, the maternal ECG represents the dominant signal
component.  Despite a strong noise component, the spike-like ventricular
complexes of the fetal signal are present. The noise floor is due to action
potentials of the muscular tissue surrounding the electrodes and other
measurement noise. (Note the small overall amplitude of the signal.)  The
nonlinear filter described in this paper can be used to extract the fetal
signal in a two-step procedure: in the first step the maternal signal is
cleaned up by considering the noise level to be of the magnitude of the fetal
spikes. This identifies the noise and the fetal component together as a
contamination of the maternal waveform.  The difference between the abdominal
recording and the clean maternal component then yields a noisy fetal
signal. The fetal ECG can be separated from the noise in a second nonlinear
noise reduction step. The result of this procedure is shown in
Fig.~\ref{fig:fekg}. The abdominal recording (upper trace) is processed in two
steps to yield a fairly clean fetal ECG (lower trace). The algorithm described
in this paper allows the extraction of the fetal ECG as a real time application
on a Laptop PC (Pentium processor at 133~MHz, Linux operating system) for a
sampling rate of 250~Hz.

In conclusion, we have demonstrated that by certain modifications to the
algorithms described in the literature, nonlinear projective noise reduction
can be turned into a signal processing tool that can in many situations run in
real time in a data stream. Necessary ingredients are (1) the formulation of
the algorithm as a causal filter, relying only on information that is available
at recording time, (2) a general speed-up of the procedure by restricting
neighbour search to the immediate past (at the expense of peak noise reduction
performance), and (3) further speed-up by using a data base of previously
processed phase space points. By putting a time restriction on the data base 
for the formation of local subspaces, processing of non-stationary signals with
slowly drifting parameters becomes also possible.

Carsten Sternickel was so kind to let us use his MCG recordings. Leci Flepp
provided the NMR laser time series and John F. Hofmeister made his fetal ECG
recordings available. We thank Daniel Kaplan, Rainer Hegger, and
Holger Kantz for useful discussions.  This work was supported by the SFB 237 of
the Deutsche Forschungsgemeinschaft.
\eject


\begin{thebibliography}{10}

\bibitem{Ablabla}
H.~D.~I. Abarbanel,
   ``Analysis of Observed Chaotic Data'',
   Springer, New York (1996).

\bibitem{KantzSchreiber}
   H. Kantz and T. Schreiber, 
   ``Nonlinear Time Series Analysis''.
   Cambridge Univ. Press, Cambridge (1997).

\bibitem{ks}  
E.~J.\ Kostelich and T.\ Schreiber,
   Phys.\ Rev.\ E {\bf 48} 1752 (1993). 

\bibitem{Davies}
M.~E. Davies,
   Physica D {\bf 79}, 174 (1994).

\bibitem{marcus} M. Richter and T. Schreiber,
   {\em Phase space embedding of electrocardiograms: Background and 
   applications}, 
   to be published (1998).

\bibitem{davies}
M.~E. Davies, 
   Int.\ J.\ Bifurcation and Chaos {\bf 3}, 113 (1992).

\bibitem{neigh}
T. Schreiber,
   Int.\ J.\ Bifurcation and Chaos {\bf 5}, 349 (1995).

\bibitem{on} P.\ Grassberger, R.\ Hegger, H.\ Kantz, C.\ Schaffrath, and
   T.\ Schreiber, 
   Chaos {\bf 3}, 127 (1993).

\bibitem{sauer} 
T. Sauer, 
   Physica D~{\bf 58}, 193 (1992).

\bibitem{chs} 
R. Cawley and G.-H. Hsu, 
   Phys.\ Rev.\ A {\bf 46} 3057 (1992);
R. Cawley and G.-H. Hsu, 
   Phys.\ Lett.\ A {\bf 166} 188 (1992).

\bibitem{processing} T. Schreiber,
   in Proceedings of ``Nonlinear Techniques in Physiological
   Time Series Analysis'', Dresden, Germany, October 1995, Springer, 
   to appear (1998).

\bibitem{takens}
   F. Takens, 
   ``Detecting Strange Attractors in Turbulence'',
   Lecture Notes in Math.\ Vol.~898, Springer, New York (1981). 

\bibitem{embed} T.\ Sauer, J.\ Yorke, and M.\ Casdagli, 
   J.\ Stat.\ Phys. {\bf 65} 579 (1991).

\bibitem{svd}
D. Broomhead and G.~P. King, 
   Physica D {\bf 20}, 217 (1986).

\bibitem{svd2}
D. Broomhead, R. Jones, and G.~P. King, 
   J.\ Phys.\ A {\bf 20}, L563 (1986).

\bibitem{vg}
R. Vautard, P. Yiou, and M. Ghil,
   Physica D {\bf 58}, 95 (1992).

\bibitem{tisean}
T. Schreiber, R. Hegger, and H. Kantz,
   {\em Practical implementation of nonlinear time series methods},
   to be published (1998); The programs can be obtained from
   {\tt http://\linebreak[1]%
      wptu38.\linebreak[1]%
      physik.\linebreak[1]%
      uni-\linebreak[1]%
      wuppertal.\linebreak[1]%
      de/\linebreak[1]%
      Chaos/\linebreak[1]%
      DOCS/\linebreak[1]%
      welcome.html}.

\bibitem{raser}
M. Finardi, L. Flepp, J. Parisi, R. Holzner, R. Badii, and E. Brun, 
   Phys.\ Rev.\ Lett.\ {\bf 68}, 2989 (1992). 

\bibitem{buzug}
H. Kantz, T. Schreiber, I. Hoffmann, T. Buzug, G. Pfister, L.~G.~Flepp, 
   J.~Simonet, R.~Badii, and E. Brun,
   Phys.\ Rev.\ E {\bf 48}, 1529 (1993).

\bibitem{SK} 
T. Schreiber and D. T. Kaplan,
   CHAOS {\bf 6}, 87--92 (1996).

\bibitem{SK2} 
T. Schreiber and D. T. Kaplan,
   Phys.\ Rev.\ E.\ {\bf 53} R4326 (1996).
   
\bibitem{marcus2} M.~Richter, T.~Schreiber, D.~T.~Kaplan,
   IEEE Trans.\ Bio-Med.\ Eng.\ {\bf 45}, 133 (1998).

\bibitem{recording}
J.~F Hofmeister, J.~C. Slocumb, L.~M. Kottmann, J.~B. Picchiottino,  and
   D.~G. Ellis,
   Biomed.\ Instr.\ Technol.\ {\bf 9}, 391 (1994).

\end{thebibliography}
\end{document}